\documentclass[english, physrev, aps,twocolumn, 10pt, amsfonts, amssymb, amsmath]{revtex4-1}

\usepackage{mathtools}

\usepackage[inline]{enumitem}
\setlist[itemize]{itemsep=0pt, leftmargin=*, widest*=0}
\setlist[enumerate]{
 itemsep=0pt,
 leftmargin=*,
 nosep,
 label={\bf\roman*})
}

\usepackage{siunitx}
\DeclareSIUnit{\Isat}{\ensuremath{\mathit{I}_\text{sat}}}
\DeclareSIUnit{\IsatR}{\ensuremath{\mathit{I}_\text{sat}^\text{R}}}
\newcommand{\SIdash}[2]{\SI[number-unit-product = -]{#1}{#2}}

\usepackage{mhchem}
\usepackage{graphicx, float, ctable}

\usepackage{xcolor}
\definecolor{darkblue}{rgb}{0,0,0.7}

\usepackage{babel}

\usepackage[normalem]{ulem}

\PassOptionsToPackage{hyphens}{url} 
\usepackage{hyperref}
\hypersetup{
 colorlinks=true,
 citecolor=darkblue,
 linkcolor=darkblue,
 urlcolor=darkblue
}
\usepackage[capitalise]{cleveref} 

\usepackage[english=american]{csquotes}

\begin{document}

\title{Shelving spectroscopy of the strontium intercombination line}

\author{I. Manai$^{1, *}$} 
\author{A. Molineri$^{2}$} 
\email[These authors contributed equally to this work.]{}
\author{C. Fr\'ejaville$^{2}$} 
\author{C. Duval$^{1}$}
\author{P. Bataille$^{1}$}
\author{R. Journet$^{2}$} 
\author{F. Wiotte$^{1}$}
\author{B. Laburthe-Tolra$^{1}$}
\author{E. Mar\'echal$^{1}$}
\author{M. Cheneau$^{2}$} 
\author{M. Robert-de-Saint-Vincent$^{1}$} 
\email[Corresponding author: ]{martin.rdsv@univ-paris13.fr}

\affiliation{$^1$ Laboratoire de Physique des Lasers, CNRS, Universit\'e Paris 13, Sorbonne Paris Cit\'e, F-93430 Villetaneuse, France \\
$^2$ Laboratoire Charles Fabry, Institut d'Optique Graduate School, CNRS, Universit\'e Paris Saclay, F-91120 Palaiseau, France}

\date{\today}


\begin{abstract}
We present a spectroscopy scheme for the 7-kHz-wide 689-nm intercombination line of strontium. We rely on shelving detection, where electrons are first excited  to a metastable state by the spectroscopy laser before their state is probed using the broad transition at 461\,nm.  As in the similar setting of calcium beam clocks, this enhances dramatically the signal strength as compared to direct saturated fluorescence or absorption spectroscopy of the narrow line. We implement shelving spectroscopy both in directed atomic beams and hot vapor cells with isotropic atomic velocities. We measure a fractional frequency instability $\sim 2 \times 10^{-12}$ at 1\,s limited by technical noise - about one order of magnitude above shot noise limitations for our experimental parameters. Our work illustrates the robustness and flexibility of a scheme that can be very easily implemented in the reference cells or ovens of most existing strontium experiments, and may find applications for low-complexity clocks.
\end{abstract}

\maketitle

\section{Introduction}
\label{sec:Intro}

Narrow optical lines of atoms or molecules have found many applications ranging across laser cooling below micro-Kelvin temperatures\,\cite{Katori1999, Curtis2001, Duarte2011, Stellmer2013}, velocity-selective probing of ultracold gases \cite{Dareau2015}, molecular photo-association spectroscopy \cite{Stwalley1999, Jones2006}, and of course as frequency references (atomic optical clocks) \cite{Bloom2014, Ludlow2015, Hollberg2005, Bize2005}. 
Exploiting the benefits of narrow resonances however imposes challenging requirements on the stability of the laser frequency. While well-established techniques exist to lock the laser frequency on an artificial reference such as an optical cavity, thereby providing short-term stability, absolute referencing with respect to the atomic or molecular line remains necessary to achieve a long-term stability.

The linewidths of the intercombination lines of alkaline-earth or alkaline-earth-like species range from tens of Hertz for magnesium \cite{Friebe2008}, to hundreds of kilo-Hertz for ytterbium. For such low values, absolute frequency referencing,  a cornerstone of atomic physics experiments, becomes a technical challenge (see e.g. \cite{Li2004} for strontium). Indeed, on the one side, the requirement for precision is increased in comparison to the routinely used dipolar allowed transitions. On the other side, the standard referencing techniques, such as saturated fluorescence or absorption spectroscopy, suffer from a drastic reduction of their signal strength.

In this paper, we demonstrate a simple spectroscopy scheme for the \SIdash{7}{\kHz}-wide intercombination line of \ce{^88Sr}. 
It consists in first exciting the intercombination line $\ce{^1S0} \rightarrow \ce{^3P1}$  using a 689 nm spectroscopy laser, and then probing the electronic state populations with a laser resonant with the broad transition  $\ce{^1S0} \rightarrow \ce{^1P1}$ at 461 nm. The spectroscopy thus inherits both the frequency resolution associated with the narrow transition, and the large signal associated with light scattering on a broad line.
Such optical ``shelving" detection is reminiscent of the approach now commonly used in ion (or atom) clocks \cite{Dehmelt1982, Ludlow2015} and molecular spectroscopy \cite{Wrachtrup1993}. Our system bears strong similarities to thermal calcium beam clocks\,\cite{Kersten1999, Oates1999, Kai-Kai2006, McFerran2009, Shang2017}. Unlike these, however, we do not implement Ramsey interferometry, which reduces the complexity of the setup.
By implementing the spectroscopy on two fully independent setups, with different atomic sources and beam geometries, we show that the scheme does not even need a directed atomic beam, unlike Ramsey schemes, so that it can be readily implemented on the atomic reference cells of most existing experiments. 
Based on our observations, we suggest a simple and compact design for a spectroscopy cell. We measure a frequency instability of $2 \times 10^{-12}$ at 1s, limited by technical noise. The expected fundamental limitations in our geometry are  $\sim 3 \times 10^{-14}$, which shows that shelving spectroscopy of the Sr intercombination line has potential application for a low complexity clock.

\section{Physical principle}
\label{sec:physicalprinciple}

Non-linear Doppler-free fluorescence or absorption spectroscopy is much more demanding with narrow lines than broad lines, because of a much weaker signal. Indeed, the number of scattered or absorbed photons contributing to the signal depends on:
\begin{enumerate}[widest*=2]
 \item 
 the number of atoms in a velocity class resonant with the laser, which, due to the Doppler effect, is proportional to the transition linewidth $\gamma$ in the limit of negligible power broadening;
 \item the scattering rate of each of these atoms, bounded from above by half the linewidth, $\gamma/2$.
\end{enumerate}
Thus, the signal strength can scale with the linewidth as fast as $\gamma^2$.
In the case of strontium, for instance, the linewidth of the $\ce{^1S0} \rightarrow \ce{^3P1}$ transition at \SI{689}{\nm} is $\gamma / 2\pi = \SI{7.4}{\kHz}$, about 4000 times smaller than the linewidth of the $\ce{^1S0} \rightarrow \ce{^1P1}$ transition at \SI{461}{\nm}, $\Gamma / 2\pi = \SI{31}{\MHz}$. The spectroscopy signal may then be reduced by a factor up to $(\Gamma /\gamma)^2 \simeq 10^{7}$ as compared to a spectroscopy of the broad line. Even under typical conditions of strong power broadening, the signal strength is still reduced by a factor larger than $\Gamma/\gamma$. Such a weak signal has been used already for accurate spectroscopy of the strontium intercombination line \cite{Ferrari2003, Courtillot2005, Hui2015}, but one must then take special care of the detection noise. Other approaches with an optical cavity around the sample are being explored \cite{Christensen2015}.

\begin{figure}[h]
 \centering
 \includegraphics[width=0.5\textwidth]{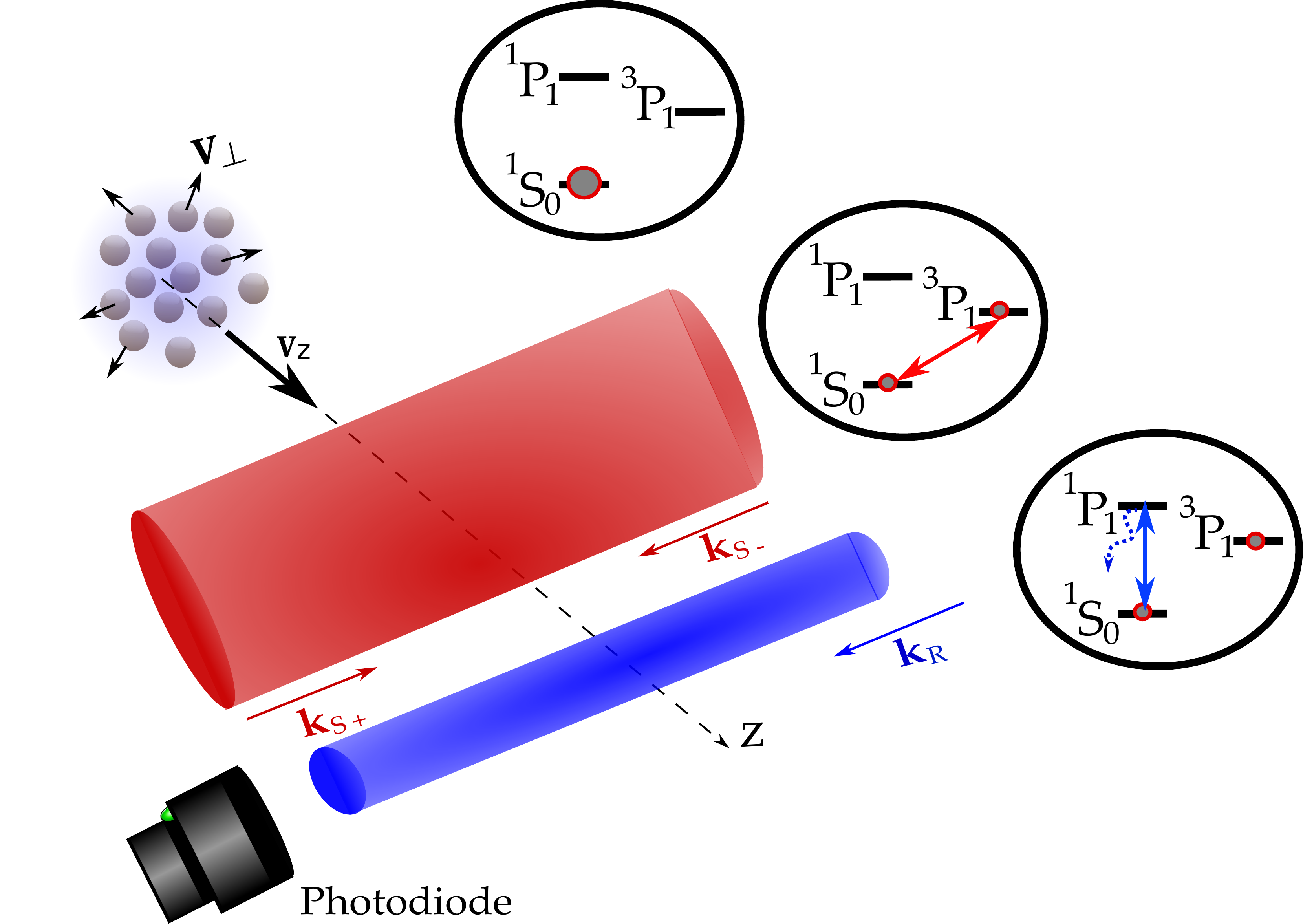}
 \caption{Shelving detection for the spectroscopy of the strontium intercombination line. Atoms first interact with the retro-reflected ``spectroscopy" laser, that probes the intercombination line. A fraction of the population is promoted to the long-lived state $\ce{^3P1}$. In a second stage, a ``readout" laser measures, by absorption, the fraction of atoms that remained in the ground state $\ce{^1S0}$. 
 }
 \label{fig:basicprinciple}
\end{figure}

The general idea of a shelving detection scheme is to increase the signal by separating the generation of the sub-Doppler feature from the detection of the electronic state populations.
To this end, we separate the atomic dynamics in two consecutive stages, which is possible thanks to the long radiative lifetime of the $\ce{^3P1}$ state (see Fig.\,\ref{fig:basicprinciple}):
\begin{enumerate}[widest*=2]
 \item the interaction with a \enquote{spectroscopy} laser driving the narrow transition, retro-reflected and saturating. 
When tuning the spectroscopy laser frequency,  the number of atoms in the excited state exhibits a sub-Doppler feature, related to the resonance of a single velocity class with both counter-propagating beams;
 \item the interaction afterwards with a non-saturating \enquote{readout} laser resonant with the broad transition, to measure the atomic ground-state population with high signal-to-noise ratio. 
The promotion of atoms to the \ce{^3P1} state by the spectroscopy laser is then detected as an increase of the readout laser transmission. 
\end{enumerate}

In doing so, we recover a factor $\Gamma/\gamma$ in the signal strength. 
It then becomes easy to create an absolute frequency referencing with the accuracy needed for applications such as laser cooling on a narrow line. For long-term laser frequency stabilization, the enhanced signal permits the use of standard detection electronics and photodiode, as well as the use of low-flux, long-lifetime atomic sources.
For the more demanding applications of frequency metrology, the relative photon shot-noise is significantly reduced, and actually typically becomes lower than atomic shot-noise, that becomes the fundamental limit to precision \cite{McFerran2010, Shang2017} (see \cref{sec:SNR}). The atomic shot noise itself can be significantly smaller than in optical lattice clocks, since thermal beam setups enable detection with a large atomic density \cite{Olson2019}.

Interestingly, in a regime where the interaction time with the spectroscopy laser is similar to the lifetime of the upper state of the narrow transition, coherences can play a significant role in the spectral features. Thermal calcium beam clocks, for example, optimize for this in implementing Ramsey interferometers. Fringe shapes and contrast then depend sensitively on the beam power and interaction time. In \cite{McFerran2009}, the authors maximized the fringe contrast by selecting a longitudinal velocity in the thermal beam. In the present work, on the contrary we take advantage of the full thermal velocity distribution to wash out the interference fringes, which ensures a strong robustness of the lineshapes against variations of the spectroscopy beam power.

\section{Experimental setups}
\label{sec:experimental_setups}

In this section, we describe the implementation of our shelving spectroscopy in two different settings, that illustrate its robustness and applicability in most existing experiments.

In the first setting, a directed thermal beam of atoms crosses consecutively the spectroscopy and the readout beams, which both propagate perpendicularly to the atomic beam and do not overlap with each other.
Most of the data presented here was obtained using this first setup.

In the second setting, a gas with nearly isotropic velocity distribution fills a hot vapor cell. The cell is traversed by the spectroscopy and readout beams, which are overlapped on the same propagation axis. To ensure that atoms interact first with the spectroscopy beam, its diameter is chosen larger than that of the readout beam.

\subsection{Atomic beam}
\label{subsec:atomic beam}

\begin{figure}[h]
 \centering
 \includegraphics[width=0.5\textwidth]{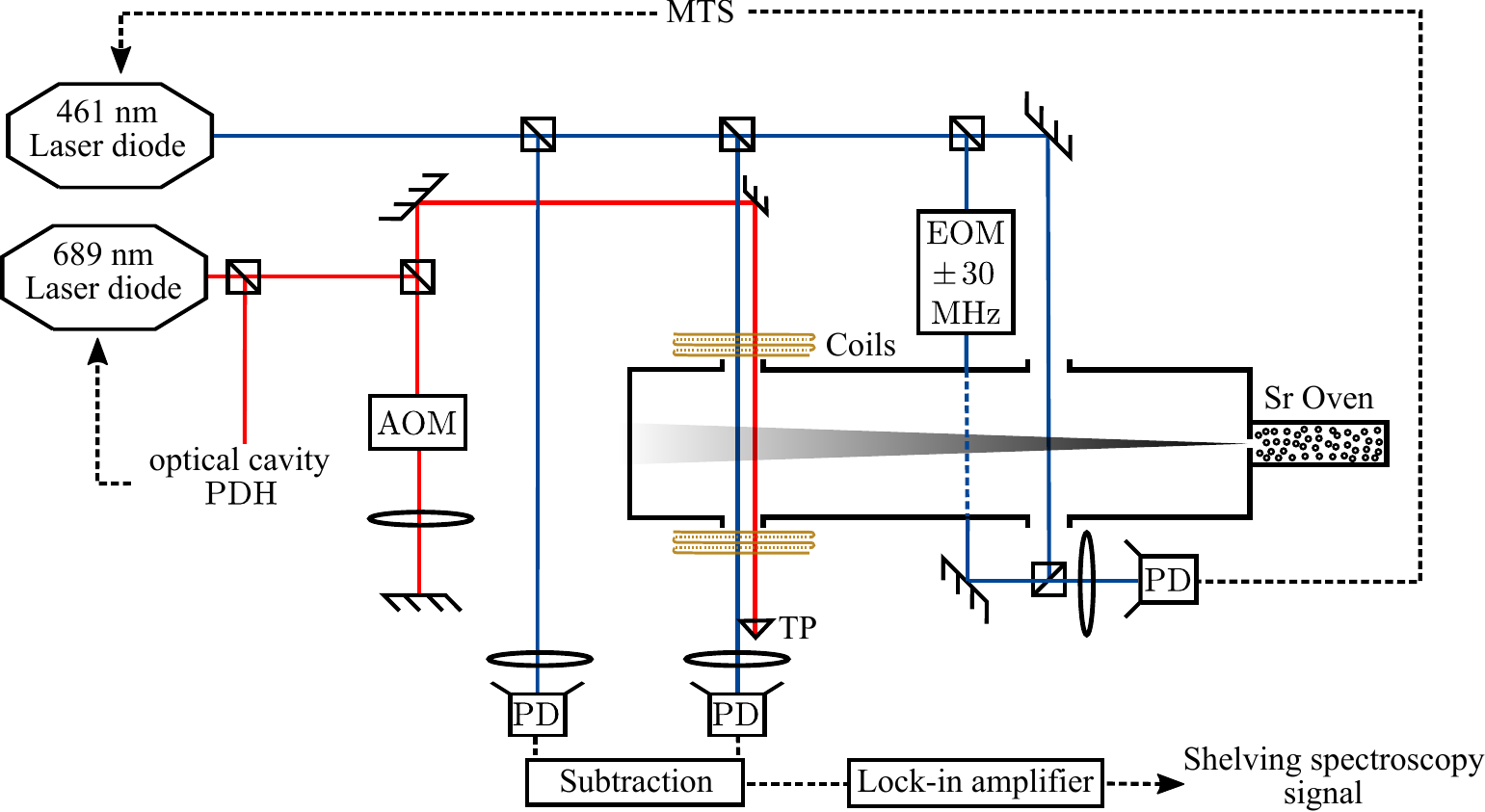}
 \caption{Overview of the atomic beam experimental setup.
  The effusive atomic beam interacts with lasers in two areas. In the first one (right), the readout laser (blue) is referenced on the broad \SIdash{461}{\nm} transition by modulation-transfer spectroscopy (MTS) for locking. The second area (left) is dedicated to the shelving spectroscopy and involves both the readout laser (blue) and the spectroscopy laser (red). 
The spectroscopy laser is frequency-modulated by means of an acousto-optic modulator (AOM) driven with sidebands. The laser carrier frequency is locked on an ultra-stable optical cavity (not shown) using the Pound-Drever-Hall (PDH) technique. A homogeneous magnetic field, generated by a pair of identical coils, splits the Zeeman levels. TP, trihedral prism; PD, photodiode.}
 \label{fig:experimental setups}
\end{figure}

In our first setup, depicted on \cref{fig:experimental setups}, a thermal beam of \ce{^88Sr} atoms is produced by an oven heated to typically \SI{480}{\celsius}. The vacuum is maintained by a \SIdash{20}{\L\per\s} ion pump.  The oven aperture (\SIdash{4}{\mm} diameter) is filled with 44 stainless steel micro-tubes (\SIdash{190}{\um} inner radius, \SIdash{15}{\mm} length) that produce an effusive directional beam\,\cite{Huckans2018}. 
The spectroscopy and readout laser beams intersect the atomic beam perpendicularly, about \SI{15}{\cm} after the oven aperture.
The (longitudinal) distribution of velocities along the atomic beam axis is $n_1(v) \propto v^2 \exp(- v^2/2mk_B T)$, with mean velocity $\overline{v}_1 = \SI{410}{\m\per\s}$. 
However, due to narrow Doppler selection by the spectroscopy laser along its axis, the atoms sensitive to it have a modified longitudinal velocity distribution: $n_2(v) \propto v \exp(- v^2/2m k_B T)$, with mean velocity $\overline{v}_2 = \SI{330}{\m\per\s}$.

The atomic beam consecutively passes through two areas. The first is dedicated to locking the readout laser on the broad line $\ce{^1S0} \rightarrow \ce{^1P1}$ by modulation transfer spectroscopy\,\cite{Camy1982}. After sufficient time for complete de-excitation, the atoms enter the second area, dedicated to the shelving spectroscopy, where they cross in close succession the spectroscopy and the readout beams. Both of these laser beams are precisely parallel to each other, and roughly perpendicular to the atomic beam. Only the spectroscopy beam is modulated in frequency in this area. 
Note that one could easily increase the number of atoms interrogated in the laser beams' volumes, and hence the signal strength, by interchanging the readout beam locking area and the shelving spectroscopy area.

Both lasers are generated by extended cavity laser diodes. 
The spectroscopy laser is frequency-stabilized to a linewidth of about \SI{1}{\kHz} by the Pound--Drever--Hall technique (PDH) \cite{Drever1983} on the TEM$_\text{00}$ mode of an ultra-stable Fabry--P\'erot cavity (Stable Laser Systems) with \SI{60}{\kHz} linewidth. 
The frequency difference between the cavity mode and the atomic resonance is controlled by two acousto-optic modulators (AOM) in double-pass configuration. 
One of these modulators, shown in Fig.\,\ref{fig:experimental setups}, is also used to modulate the spectroscopy light frequency
\footnote{The radio-frequency driver for this AOM is an AD9852 digital synthesizer. We use the FSK function and define the frequency modulation as follows: a quarter of the period is a linear chirp increasing the frequency, a quarter period at constant frequency, a quarter period linear chirp to low frequency, a quarter period at constant frequency. The discretization in instantaneous frequency of the RF synthesis is kept below 0.6 kHz, much lower than the modulation amplitude and than $\gamma$.}.

In the shelving spectroscopy area, the spectroscopy beam has a \SIdash{5.4}{\mm} waist, and it is truncated to a diameter $D = 10\,$mm.
The corresponding characteristic interaction time with the atoms, or transit time, is $D / \overline v_2 = 30\,\rm{\mu}$s. The beam is retro-reflected by a trihedral prism to ensure minimal deviation from parallelism, in the \SI{10}{\micro\radian} range, a critical point to avoid biases and broadening from first-order Doppler effect -- see \cref{subsec:Doppler_shifts}. The retro-reflected beam has a small lateral displacement, $\lesssim \SI{1}{\mm}$, along the atomic flux.
The power of the incident spectroscopy beam is varied from \SIrange{26}{470}{\micro\W}. This corresponds to a saturation parameter $I/ I_{\rm{sat}}$ between \numlist{23;400}, or Rabi frequencies $\Omega/2\pi$ between \SIlist{26;110}{\kHz}.
Here, $I$ is the peak intensity of the incident beam on its axis inside the vacuum chamber, and $I_{\rm{sat}} \approx \SI{3}{\micro\W\per\cm^2}$ is the saturation intensity of the $\ce{^1S0} \rightarrow \ce{^3P1}$ transition. The retro-reflected beam is weaker by a factor 0.85 due to the uncoated viewports.

The propagation axis of the readout beam is set \SI{2}{\mm} after the truncated edge of the spectroscopy beam. The mean travel time between the spectroscopy and readout beams is $\approx \SI{6}{\us}$, smaller than the \SIdash{21}{\us} lifetime of the \ce{^3P1} state.
The readout beam is elliptical with waists $\SI{440}{\um}$ and $\SI{900}{\um}$, respectively along the atomic beam and orthogonal to it. 
Its power is set to \SI{66}{\micro\W}, which corresponds to a saturation parameter of about \num{0.25}.
The absorption by the atomic beam is measured by an amplified photodiode 
and is typically \SI{7}{\percent}.
Another photodiode monitors the readout beam power upstream of the atomic beam; the two photodiode signals are subtracted to cancel the effects of intensity fluctuations in the readout beam. The resulting signal is amplified and demodulated by a lock-in amplifier synchronized with the modulation of the spectroscopy AOM frequency. 

In the shelving spectroscopy area, a uniform magnetic field of \SI{3.4}{G} is generated by a pair of coils along the light propagation axis to separate the Zeeman levels of the \ce{^3P1} state (with Land\'e factor \SI{2.1}{\MHz\per G}). The polarization of the spectroscopy light is linear and perpendicular to the magnetic field, such that the $\pi$ transition $\ce{^1S0}(m_j = 0) \rightarrow \ce{^3P1}(m_j = 0)$ is forbidden. A cross-over line between the $\sigma^+$ and $\sigma^-$ transitions nevertheless provides a signal effectively insensitive to magnetic fields in the Gauss range.

\subsection{Hot vapor cell}
\label{subsec:hot vapor cell}

Our second setup is a hot vapor cell, in which we demonstrate the shelving spectroscopy despite the absence of a well directed atomic flux. 
The expected advantage of the cell geometry is to probe a much denser ensemble of atoms, and thus to obtain a large signal strength even for modest gas temperatures.

The cell consists of a \SIdash{60}{\cm} long, stainless steal tube connecting two uncoated BK7 viewports. A crucible welded in the middle serves as a strontium reservoir. The tube's internal and external diameters are \SIlist{8;14}{\mm}, respectively.
A welded nipple located close to one of the viewports provides a connection to the exterior through an angle valve. The vacuum pressure inside the cell is brought below $ \SI{e-4}{\milli bar}$ using a turbomolecular pumping station. Once under vacuum, the cell can be isolated by closing the valve and the pumping station can be removed, as there is no need for continuous pumping at this level of pressure. We verified that such moderate vacuum pressure is sufficient to avoid the collisional broadening of the narrow intercombination line (see \cref{subsec:pressure}).

The crucible is a hollow cylinder whose inner volume has a diameter of \SI{16}{\mm} and a height of \SI{50}{mm}. It is filled with several grams of strontium. A coaxial heating cable is wound around the crucible, and raises the temperature to about \SI{390}{\celsius}, where the strontium vapor pressure is around \SI{e-4}{\milli bar} \cite{Rumble2018}.
Thermal insulation is ensured by wrapping the heated region in ceramic wool and aluminum foil. In practice, the temperature of the steel tube away from the heated region quickly drops, and there is no need for active cooling of the cell's ends. The length of the cell tube prevents strontium deposition on the ambient-temperature viewports.

The spectroscopy and readout beams are generated by extended-cavity diode lasers, locked on fixed frequency references in a very similar way as described for the first setup. The most notable differences are:
\begin{enumerate}[widest*=3]
 \item a hollow cathode lamp serving as a reference 
to stabilize the readout laser frequency on the $\ce{^1S0} \rightarrow \ce{^1P1}$ transition;
 \item the larger linewidth  of the Fabry--P\'erot cavity (\SI{200}{\kHz}), used to stabilize the frequency of the spectroscopy laser to better than \SI{1}{\kHz} with the PDH technique;
 \item the power amplification of the spectroscopy laser with a master-slave system. The master laser is locked on the optical cavity, and the frequency modulation for the spectroscopy is applied by a double-pass AOM on the master, before injection of the slave laser. 
\end{enumerate}

The spectroscopy and readout beams are superimposed on the center axis of the vapor cell using a low-pass dichroic mirror. The spectroscopy beam initially has a Gaussian profile with a waist of \SI{4}{\mm} and is clipped by the \SIdash{8}{\mm} diameter tube. It is retro-reflected by a trihedral prism. The light intensities used here are in the same range as in our first setup. The readout beam has a Gaussian profile of waist \SI{400}{\um} and its power is set to \SI{30}{\micro\W}, corresponding to $I/I_{\rm{sat}}^{\rm{R}} \simeq$\,\num{0.15} ($I_{\rm{sat}}^{\rm{R}} = 42$\,mW/cm$^2$ for the readout transition). 
The absorption of this beam by the atomic vapor cell is typically between \SIlist{50;80}{\percent}, depending on the exact temperature of the crucible. This absorption rate is 10 times higher than in the first setup, providing in principle a gain in the spectroscopy signal amplitude of order 5 (see Appendix \ref{sec:SNR}) while using a lower temperature.
The spectroscopy and readout beams are separated by a dichroic mirror after passing through the cell. The spectroscopy beam continues towards the retro-reflector, while the readout beam is directed towards a non-amplified photodiode to measure the absorption signal.
A laser-line band-pass filter (center at \SI{460}{\nm}, \SIdash{10}{\nm} bandwidth) is placed in front of the detector to block residual light from the spectroscopy beam with an optical density of 6.
The absorption signal is amplified and demodulated by a lock-in amplifier. In contrast to the first setup, we do not subtract a measure of the incident readout power from this signal.

The typical transit time of the atoms contributing to the spectroscopy is shorter in this setup. First, by Doppler selection, all the atoms that contribute to the sub-Doppler signal travel orthogonally to the beams. Due to the colinear propagation of the spectroscopy and readout beams, the distance over which the atoms interact with the spectroscopy beam is limited by its radius, 4\,mm, instead of its diameter.
Second, the atoms stop interacting with the spectroscopy beams as soon as the scattering of photons from the readout beam dominates their internal dynamics. We estimate that this occurs 1.6 mm away from the beam axis, where the scattering rate from the readout laser equals the Rabi coupling frequency from the spectroscopy laser. This leads to a typical travel distance in the spectroscopy beam $D \approx \SI{2.4}{\mm}$, compared to \SI{1}{\cm} in the atomic beam setup. 
From any point on the cell walls, the velocity distribution of atoms intersecting the readout beam with vanishing Doppler shift is $n_2(v)$ (see \ref{subsec:atomic beam}), with mean velocity $\overline{v_2} = \SI{310}{\m\per\s}$.
From this we deduce a typical transit time $\sim 8\,\mu$s, smaller than the $\ce{^3P1}$ state lifetime. Coherences on the intercombination transition are thus well preserved during the interrogation. 
However, simple numerical simulations based on the optical Bloch equations show that averaging over the velocity distribution washes out the spectral signatures of coherence, which ensures robust spectral shapes - as supported by our observations.

The current circulating in the heating cable produces a magnetic field of about \SI{2.3}{G}, oriented perpendicular to the beam propagation axis. The polarization of this beam being mixed in the experiment, we are able to see the three transitions associated with the Zeeman sublevels of the \ce{^3P1} state, plus two cross-over lines.  The cross-over line between the two $\sigma$ lines is superimposed on the $\pi$ line. In the following we consider only this central spectral line, which is insensitive to the magnetic field.

\begin{figure}[th]
 \centering
 \includegraphics[width=0.5\textwidth]{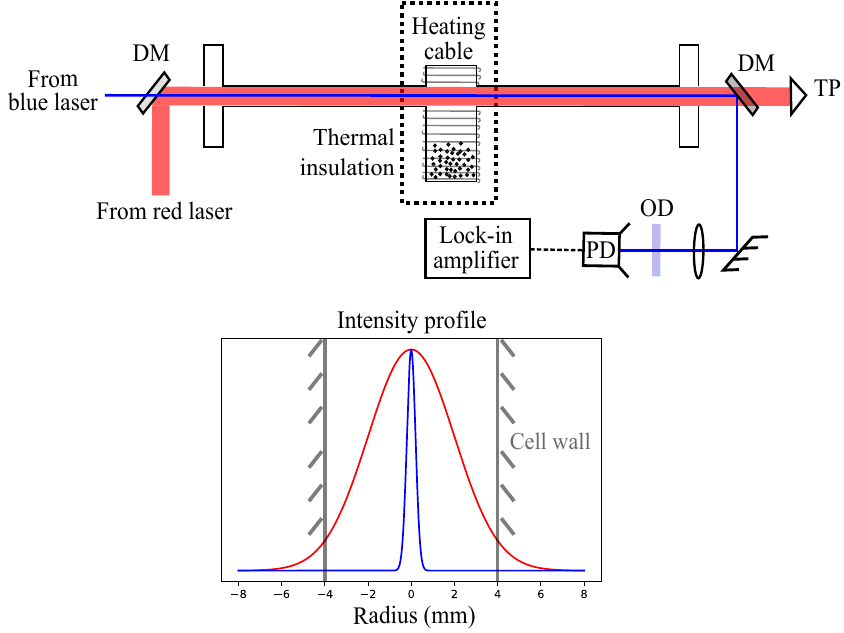}
 \caption{Overview of the hot vapor cell experimental setup. The heating cable wound around the strontium reservoir generates a magnetic field oriented perpendicularly to the laser beam propagation axis. 
The spectroscopy laser is locked on a Fabry-P\'erot cavity (not shown), the readout laser on a hollow cathod lamp (not shown). The spectroscopy laser is frequency-modulated using an AOM.
Bottom panel: intensity profiles of the spectroscopy beam (red line) and readout beam (blue line), normalized to one at the center, along a diameter of the cell tube. DM, Dichroic Mirror; TP, Trihedal retro-reflection Prism; PD, photodiode; OD, laser-line filter (optical density of 6 at 689\,nm).}
 \label{fig:experimental setups LCF}
\end{figure}

\section{Typical spectra}
\label{sec:typical_spectra}

\begin{figure}
 \centering
 \includegraphics{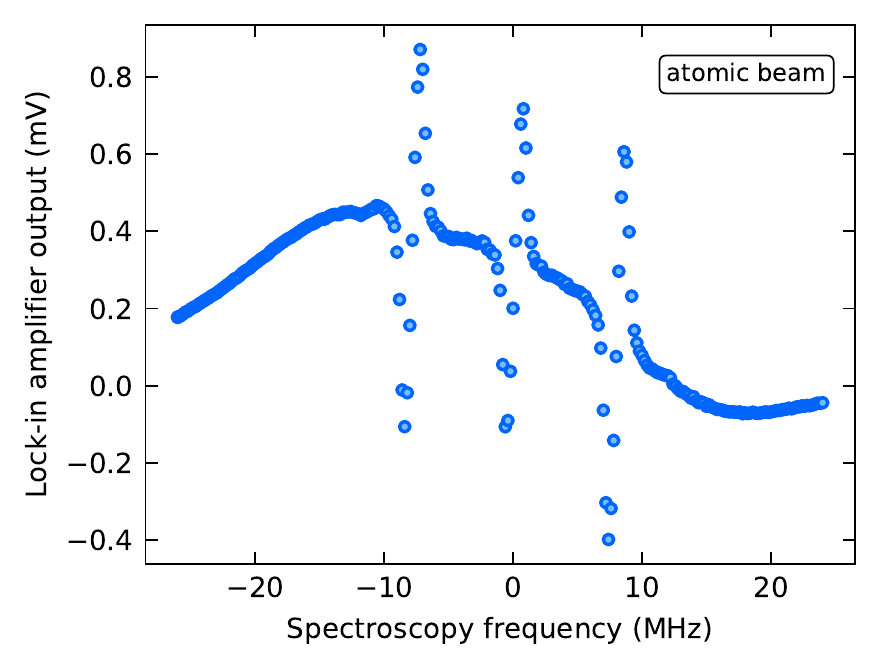}
 \caption{
  Broad view of the spectrum, with a strong modulation of the spectroscopy beam frequency (\SI{2}{\MHz} peak-to-peak) to inflate the wide Doppler feature.
  The latter originates from the interaction of the spectroscopy beam with all atoms with velocities such that $\vec k_{\rm{R}} \cdot  \vec v < \Gamma$, that are probed by the readout beam.
  The three sub-Doppler resonances correspond to the $\sigma^\pm$ transitions and their cross-over line. They are split by \SI{7.1}{\MHz} thanks to a \SIdash{3.4}{G} magnetic field.
  The origin of the frequency axis is arbitrarily set to the center of the cross-over line.
 }
 \label{fig:broad_spectrum}
\end{figure}

We now present typical spectra obtained with both our setups.
\cref{fig:broad_spectrum} first shows a broad view of the spectroscopic signal using the atomic beam setup. In order to reveal the broadest spectral features, the modulation amplitude of the spectroscopy beam frequency is chosen very large (\SI{2}{\MHz} peak-to-peak) and dominates over all other broadening mechanisms. The modulation frequency is set to \SI{10}{\kHz}.  This spectrum is acquired in ascending frequency order, with 100\,ms lock-in amplifier integration time and 500\,ms delay between consecutive measurement points.

The wide dispersive feature spanning the whole frequency range originates from the interaction of the spectroscopy laser with all atoms that can absorb light from the readout beam. These atoms have velocities such that $\vec k_{\rm{R}} \cdot \vec v \leq \Gamma$, where $\vec k_{\rm{R}}$ is the wavevector of the readout beam. This corresponds to Doppler shifts for the spectroscopy laser of up to $\Gamma \cdot k_{\rm{S}} / k_{\rm{R}} \simeq 2\pi \times \SI{21}{\MHz}$  ($k_{\rm{R}}$ and $k_{\rm{S}}$ are the norms of the readout and spectroscopy beam wavevectors, respectively).
On top of this background, three Doppler-free resonances emerge, due to the interaction of the spectroscopy beam and its retro-reflection with atoms belonging to the same velocity class.
The outer lines correspond to the $\sigma$-transitions, and the central line is the cross-over between the two, which is insensitive to the magnetic field to first order.

\begin{figure}
 \centering
 \includegraphics{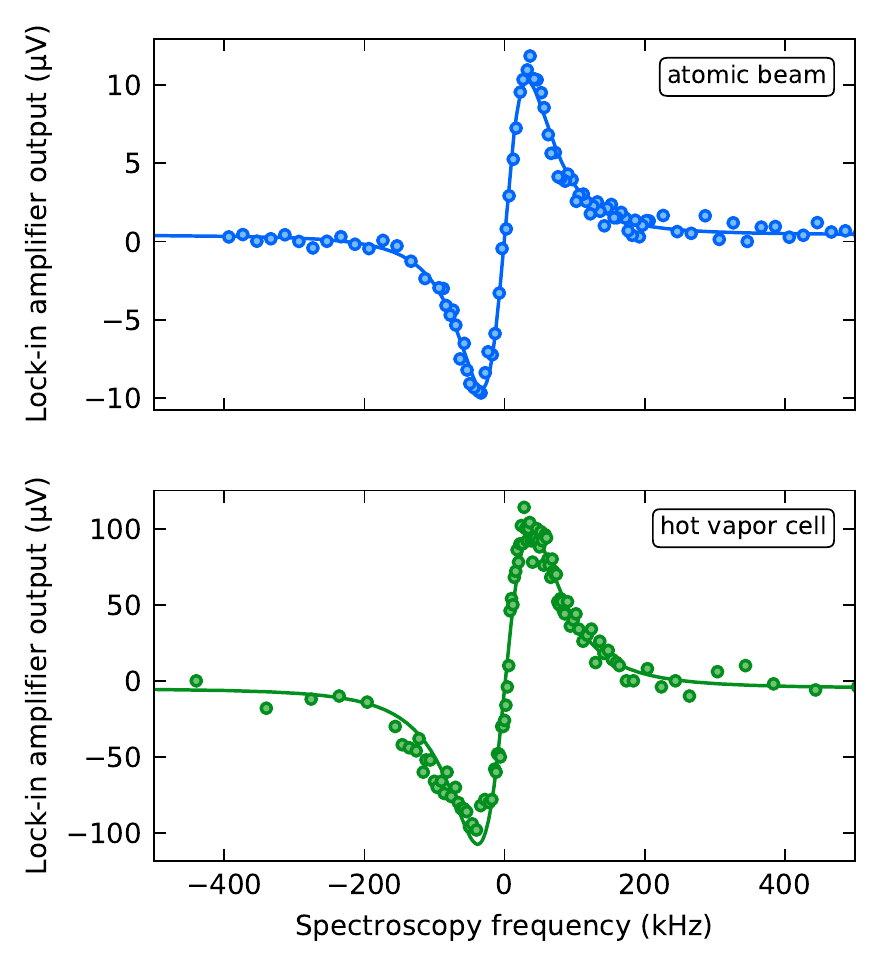}
 \caption{
  Dispersion profile from the $\sigma^\pm$ cross-over line and $\pi$ line, in the conditions of section \ref{sec:typical_spectra}. \textbf{Top panel:} measurement performed on the atomic beam. \textbf{Bottom panel:} measurement performed on the hot vapor cell. 
  In both panels the circles represent the measured data and the solid lines represent the best fits with the derivative of a Lorentzian profile, plus a constant offset baseline. The origin of the frequency axis is arbitrarily set to the center of the line. 
  The key parameter in this line profile is the spectroscopy beam intensity. It is set to $I \simeq 83 I_\text{sat}$ for both measurements.
  The full widths at half maximum derived from the fits are \SI{114}{\kHz} (atomic beam) and \SI{133}{\kHz} (hot vapor cell).
 }
 \label{fig:spectra}
\end{figure}

We now zoom in on the central Doppler-free resonance, for both setups.
In \cref{fig:spectra}, we show spectra obtained with the experimental parameters which, on the atomic beam setup, optimize the uncertainty of the measurement of the line center position (see \cref{sec:precision}).
In both setups, the peak intensity of the incident spectroscopy beam is $I \simeq 83 I_\text{sat}$ and its frequency is modulated at \SI{2.5}{\kHz}, slow enough to ensure that the atomic response is quasi-static, with no dephasing
\footnote{We also tried to modulate at larger frequencies, as shown on Fig.\,\ref{fig:broad_spectrum}. At \SI{10}{\kHz}, one quadrature displays a dispersive-like signal, suited for the spectroscopy, while the other quadrature has a Lorentzian shape that resembles a time-averaged, low-noise shelved population measurement. However, these signals are extremely sensitive to the demodulation phase, which can cause drifting offsets and asymmetric line shapes. This sensitivity worsens as the the modulation frequency is increased.}. 
The modulation amplitude is \SI{66}{\kHz} peak-to-peak in the atomic beam setup and \SI{22}{\kHz} in the hot vapor cell.
The lock-in amplifier integration time is \SI{1}{\s} (resp. \SI{3}{\s}), and, in order to prevent spectrum distortion, a new point is acquired every \SI{5}{\s} (resp. \SI{10}{\s}) in order of ascending frequency. The absorption on the readout beam is around \SI{7}{\percent} (resp. \SI{80}{\percent}) when the spectroscopy beam is off-resonant.

The solid lines in \cref{fig:spectra} show the best fits to our data using the derivative of a Lorentzian profile, plus an offset. We find a full width at half maximum (FWHM) of $\SI{114}{\kHz}$ (resp. \SI{133}{\kHz}), resulting from the addition of several broadening mechanisms with similar contributions: the finite transit times, the spectroscopy beam intensity, and the finite modulation amplitude; see \cref{sec:precision}.
The fit uncertainty on the line center is close to \SI{500}{\Hz} in both setups.

\section{Frequency measurement performances}
\label{sec:precision}

Close to the resonance, in the linear part of the spectrum, the spectroscopy signal provides a measure of the frequency difference between the laser and the atomic resonance. However, even with an ideal laser source displaying a perfectly stable frequency, the measurement process itself is affected by noise. The natural short-term timescale of the measurement being the lock-in amplifier integration time, we estimate the instability of the frequency measurement at this time scale (abbreviated frequency instability) from the ratio between the noise of the lock-in amplifier signal and the slope in the linear part.
Below, we will discuss these two aspects consecutively, based on data from the atomic beam setup. We will then compare this frequency instability at 1\,s to the fit uncertainty on line centers from entire spectral datasets. The fit uncertainties are sensitive to the line shapes, and to baseline drifts appearing in some parameter ranges.

\subsection{Signal slope optimization}
\label{subsec:slope}

Although the spectroscopy signal broadens with both the intensity of the spectroscopy beam and the amplitude of the frequency modulation for the lock-in amplifier, the slope of the signal best characterizes the figure of merit of the spectroscopy. Here we show how both parameters above optimize the slope.
The modulation frequency remains $f_{\rm{mod}} = $2.5\,kHz. 

\begin{figure}
 \centering
 \includegraphics{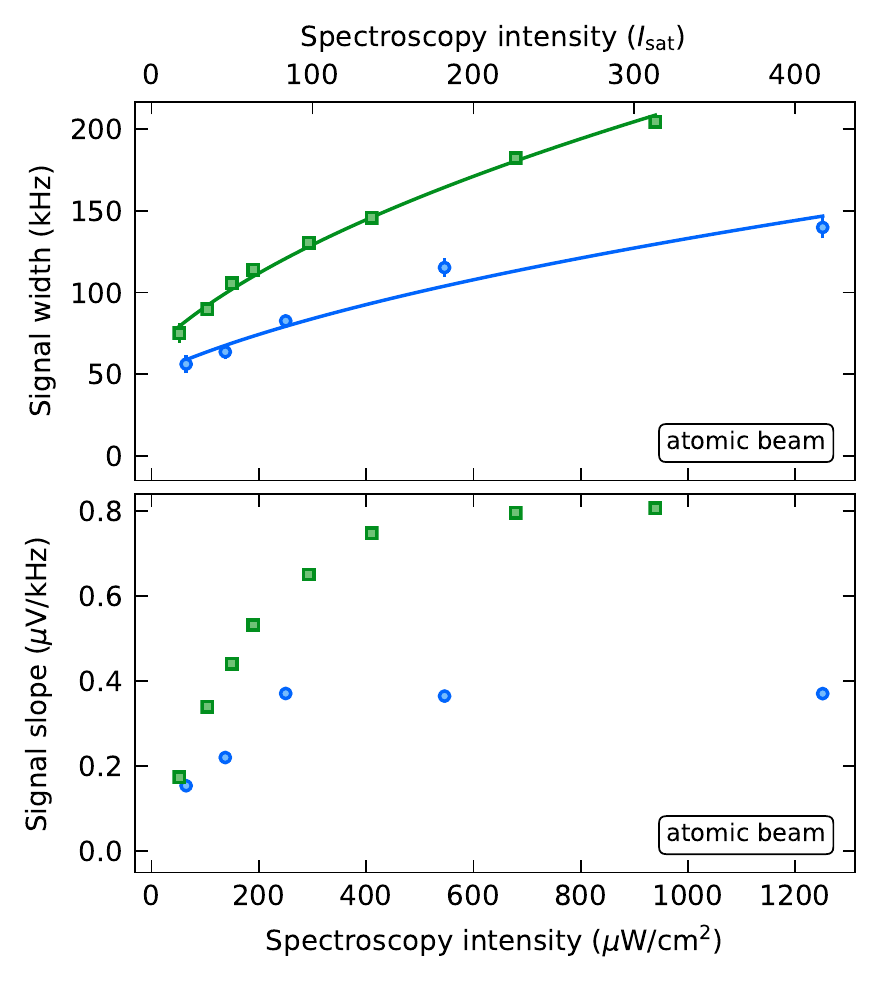}
 \caption{
  \textbf{Top panel:} fitted Lorentzian FWHM as a function of the incident spectroscopy beam intensity. 
  The data marked with blue circles was recorded with a fixed modulation  amplitude of \SI{25}{\kHz} peak-to-peak. The blue solid line shows the best fit using the empirical function $\gamma(I)/2\pi = \sqrt{a^2 I + b^2}$.
  The data marked with green squares was recorded with a peak-to-peak frequency-modulation amplitude scaled to $0.8\times(\gamma(I)/2\pi)$, to maximize the signal slope (see text). As a guide to the eye, the latter data is fitted with the same model function, shown as green line. The error bars on the data denote the spectrum fit uncertainties.
     \textbf{Bottom panel:} slope of the signal as a function of the spectroscopy beam intensity. The blue circles and green squares are derived from the same measurements as in the top panel.
 }
 \label{fig:width/slope}
\end{figure}

First, we begin by investigating the dependency of the spectral width on the incident spectroscopy beam intensity, see \cref{fig:width/slope}, top panel.
The blue circles represent the fitted Lorentzian FWHM measured using a weak modulation amplitude of \SI{25}{\kHz} peak-to-peak.
The blue solid line shows the best fit with the empirical function $\gamma(I)/2 \pi =  \sqrt{a^2 I + b^2}$. We find $a = \SI{6.8 \pm 0.3}{\kHz \per \sqrt{\Isat}}$ and $b = \SI{50 \pm 8}{\kHz}$.  
The value $\gamma(0)/2 \pi = b$ should be dominated by the transit time broadening \cite{Borde1976}. 
We estimate that the amplitude of the frequency modulation also contributes to about \SI{10}{\kHz} to the spectral width.
The blue circles in the bottom panel of \cref{fig:width/slope} show the resulting slope in the linear part of the spectroscopy signal as a function of the spectroscopy intensity. The slope first improves with spectroscopy intensity, as the dispersive signal strength increases until power broadening overcomes transit time broadening.
At higher intensities, we would expect the increasing and dominant power broadening to ultimately deteriorate the dispersive signal slope (see Appendix \ref{sec:SNR}) -- although this trend is not clearly visible in our data.

Second, we discuss the optimal frequency modulation amplitude. At the spectroscopy intensity $I = \SI{46}{\Isat}$, we maximized the signal slope as a function of modulation amplitude, and found an optimum, corresponding to \num{0.8} times the spectral FWHM measured at low modulation. 
We argue, as detailed in \cref{sec:SNR}, that as pump power is varied, the modulation amplitude that maximizes the slope scales with the power-broadened width. 
Thus, in a second set of experiments, we repeated our measurements with a modulation amplitude scaled to always match a value equal to 0.8 times the linewidth $\gamma(I)/2\pi$ measured previously at low modulation amplitude. The resulting widths and slopes are shown in the two panels of \cref{fig:width/slope} as green squares. One sees that, although the line is broadened by about \SI{30}{\percent}, the slope can be increased by up to a factor 2 as compared to the previous data, and its dependency with spectroscopy power flattens at high powers, as expected (Appendix \ref{sec:SNR}).

\subsection{Instability and frequency measurement uncertainty}
\label{subsec:noise}

In this section, we characterize our spectroscopy signal in terms of frequency instability, and find a $2\times10^{-12}$ fractional instability in 1s. This provides an extremely good frequency reference for most cold atom experiments.
However, we also point out long-term drifts of the output of the lock-in amplifier, which would impact long term measurements. These seem related to stray spectroscopy light on the readout photodiode, that could be eliminated for clock applications.

We estimate the noise on the signal by repeated measurements with all parameters constant. The spectroscopy frequency is set away from the sub-Doppler line. 
For low spectroscopy intensity, up to $\leq$ \SI{100}{\Isat}, the noise appears uncorrelated, random from shot to shot. Its root-mean-square (r.m.s.) value is proportional to the readout beam power (as the sub-Doppler signal strength), and is reduced by subtracting a reference to the photodiode signal (see Fig.\,\ref{fig:experimental setups}). For the selected readout beam power (\SI{66}{\micro\W}), we find an r.m.s. noise value of \SI{0.5}{\uV}, independent of the spectroscopy intensity. 
Under those conditions, the signal-to-noise ratio is $\sim 20$. This value is about 10 times lower than the limit expected from fundamental shot noises, in which atomic shot noise dominates over photonic shot noise (see Appendix \ref{sec:SNR}).

Offset drifts, relevant at the minute timescale, appear at large spectroscopy intensities above \SI{100}{\Isat}. The strongest observed drifts are $\sim 0.01 \,\rm{\mu}$V/s.  
We suspect stray spectroscopy light reaching the detector to be responsible for them. Residual amplitude modulation applied to the spectroscopy beam by the AOM may then be detected by the lock-in amplifier. 
A demodulation at thrice the modulation frequency \cite{Wallard1972} could suppress this effect.

\begin{figure}
 \centering
 \includegraphics{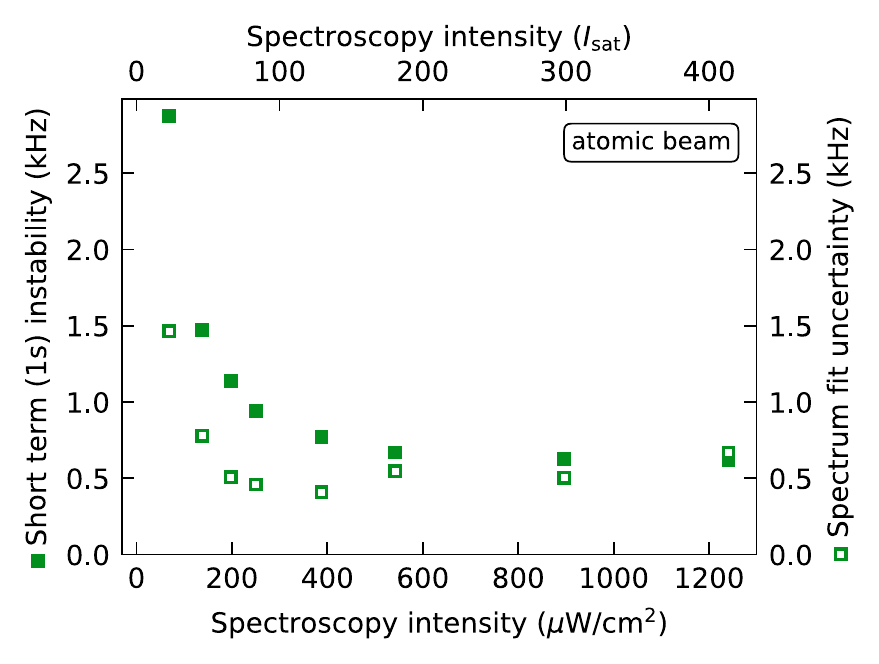}
 \caption{
Frequency instability at 1\,s (filled squares), and spectrum fit uncertainty on the line center (empty squares), as a function of spectroscopy beam intensity.  Both quantities are extracted from the same data sets, 90-shot spectra, 450\,s total acquisition time, one of which is presented in Fig.\,\ref{fig:spectra}a.
The spectra are fitted with a Lorentzian derivative, plus a baseline modeled by an offset drifting linearly with measurement time. Above 400 $\mu$W/cm$^2$, the fit uncertainty deteriorates, as baseline drifts and deviations from the assumed lineshape become significant (see text). 
 }
 \label{fig:uncertainty}
\end{figure}

We now summarize in \cref{fig:uncertainty} the performances of our spectroscopy as a function of the spectroscopy beam power, combining the effects of short-term noise (instability), long-term noise (drifts), and lineshape. This figure relies on the same spectra as in \cref{fig:width/slope} (data with scaled modulation amplitude). 

First, we estimate the short-term (1\,s) frequency instability of the measurement, realized with a 1\,s lock-in amplifier integration time, 
from the ratio of the uncorrelated noise to the dispersive signal slope. It is shown in \Cref{fig:uncertainty} (full squares) as a function of the spectroscopy beam intensity. As the uncorrelated noise is independent of the spectroscopy intensity, the frequency instability at 1\,s is simply proportional to the inverse of the slope shown in Fig.\,\ref{fig:width/slope}b (green squares). We observe that it improves quickly as the intensity is increased, and then levels off to about \SI{630}{\Hz} above  \SI{200}{\Isat}, when the power broadening starts to dominate over the transit time broadening (see \cref{sec:SNR} for a simple model). 

Second, we investigate the uncertainty of long term measurements by fitting entire spectra, made of 90 shots acquired point-by-point over 450\,s.
Since the spectroscopy laser is referenced to an optical cavity, we are then measuring the center of the line relative to that cavity (see \cref{subsec:reproducibility}). We assume Lorentzian derivative lineshapes, and include baseline drifts in the fit model function with a linear dependence on acquisition time. The uncertainty is affected by short and long-term noise, and also by deviations of the actual lineshape from the fit model function. 
The fit uncertainties on the line centers are shown as empty squares in \cref{fig:uncertainty}. They are optimal for a spectroscopy beam intensity around \SI{100}{\Isat}, reaching about \SI{450}{\Hz}. The minimal uncertainty in this intensity range results from a trade-off between several effects: a high intensity is favorable to increase the slope (see Fig.\,\ref{fig:width/slope}), but a low intensity causes smaller offset drifts and less deviations from the assumed Lorentzian lineshape.
On our setup, thus, the regime $I \simeq 100 \,I_{\rm{sat}}$ is well suited for long-term (450\,s) measurements, in particular as it reduces the drifts to values such that fitting with or without them provide the same line center measurement.
In this intensity range, the instability at \SI{1}{\s} is below 1\,kHz, i.e. a relative instability $\sim 2. 10^{-12}$.
Up to \SI{100}{\Isat}, the fact that the fit uncertainty is only about half the instability at 1\,s is statistically explained by the relatively sparse sampling of the central slope in the spectra, with only up to 10 points\, 
\footnote{For the data with $I \leq 100 I_{\rm{sat}}$, the fit uncertainties are only 15$\,\%$ larger than what would result from true Lorentzian signals with same widths, amplitudes, and noise standard deviation, given our sampling of the spectroscopy frequencies.
}.

~\\To conclude this section, we measure a relative instability at 1s of $2. 10^{-12}$. The signal-to-noise ratio (SNR) is of order 20 and  is limited by technical noise. Ultimately, if the noise level were set by the fundamental shot-noise limit (combining atomic and photonic shot noise), we  expect that the SNR could reach 220 in the same conditions. An additional improvement, by a factor of order 5 in signal, could be achieved simply by increasing the readout beam absorption, which is only 7$\%$ for these data (see \cref{sec:SNR}).

\section{Shelving detection in the hot vapor cell}
\label{sec:hotcellresults}
We now present the application of shelving detection to the hot vapor cell. We will discuss the observed performances and highlight three specific advantages of such a configuration: compactness, low complexity, and long lifetime operation. The key specificity of the hot vapor cell is that atoms are probed where the density is the highest, which enables low-temperature operation with high signal.

\subsection{Performance overview}

In Fig.\,\ref{fig:spectra}, we have 
compared spectra obtained with the atomic beam and the hot vapor cell for the same spectroscopy beam intensity of 83 $I_{\rm{sat}}$. A slightly larger linewidth (133 kHz compared to 114 kHz) is observed in the hot vapor cell, which may be attributed to a larger transit broadening given the shorter interaction length in this geometry
(see section \ref{subsec:hot vapor cell}). 
The spectrum baseline is distorted for intensities larger than \SI{100}{\Isat}. In contrast to the atomic beam setup, however, the distortions are reproducible and frequency-dependent. We attribute these effects to reflections of the spectroscopy beams off the uncoated viewports, which give rise to “ghost” signals superimposed on the main spectroscopy signal. The reflections are probably also responsible for deviations from the Lorentzian lineshape, such as the shoulder on the left flank on the spectrum shown in the bottom panel of \cref{fig:spectra}). These parasitic signals are currently a significant limitation for the long-term fit uncertainty in the hot vapor cell. They could be suppressed by using a cell with tilted viewports, or a refined lineshape model.

The SNR depends directly on the atomic density, which sets the strength of the absorption signal and also of atomic shot noise (see \cref{sec:SNR} for the fundamental limitations).
The hot vapor cell geometry is attractive in this regard, since the laser beams probe the region of the vacuum chamber with the highest atomic density, unlike in the atomic beam setup.
We typically achieve ten times stronger absorption coefficients, with 	a temperatures of  \SI{390}{\celsius} instead of  \SI{480}{\celsius}  in the atomic beam setup. 
This is a great advantage both for increasing the SNR and for the lifetime of the strontium source.

Despite the higher absorption, the SNR in the spectra acquired with our hot vapor cell remain similar to that of the atomic beam spectra. We find a frequency instability of \SI{2.0}{\kHz} for a \SIdash{3}{\s} integration time using the slope of the dispersive profile and the r.m.s. fit residuals as a noise estimate. This scales to a relative instability of $8\times 10^{-12}$ at \SI{1}{\s} ($4.6\times 10^{-12}$ at \SI{3}{\s}).
This comparatively low performance is due in part to the lineshape distortion, which artificially increase the r.m.s. fit residuals, and hence the noise estimate. Other sources also contribute to degrade the performance of our measurement, such as the small modulation amplitude (see the discussion in \cref{subsec:slope}), the electronic detection noise, the stronger transit time broadening and the smaller readout beam waist.

\subsection{Effects of pressure}
\label{subsec:pressure}

The pressure in the hot vapor cell can affect the line center position, its width and its amplitude \cite{Crane1994, Ferrari2003, Li2004}. 
By adding a controlled amount of argon buffer gas, we were able to verify that these effects are negligible as long as the \textit{total} pressure remains below \SI{e-3}{\milli\bar}, measured close to one of the viewports. Above this value, we observed a quick reduction of the signal amplitude. We did not observe a noticeable change of the position or the width of the signal before it completely disappeared. This remains consistent with \cite{Crane1994}, from which one would expect a collision broadening of 30 kHz at $10^{-3}$\,mbar of argon, small compared to our measured width.
This tolerance of the spectroscopy to large pressures is a significant advantage, as it allows the spectroscopy cell to be operated without pump within large timespans, thus simplifying the setup.

\subsection{Design outlooks}
\label{subsec:designimprobements}

In future designs, increasing the tube diameter would improve over two limitations. i) It would enable the use of a larger spectroscopy beam, to reduce transit broadening well below power broadening in the intensity range from \SIrange{50}{100}{\Isat}, where baseline drifts are no limitation. ii) The readout beam could also be larger to probe more atoms. This would increase the signal at fixed readout intensity, and simultaneously reduce the fundamental limitations to frequency instability associated with atomic shot noise (see Appendix \ref{sec:SNR}). 

However, a diameter larger by a factor of two would also increase the deposition rate on the viewports by a factor of $\sim 4$, in the ballistic regime. For our cell length, we estimate that within a few months (full-time operation) a micrometer-thick deposit may then opacify the viewports \cite{Huckans2018}. The problem could be circumvented by a buffer gas, e.g. a few $10^{-4}$\,mbar of argon, bringing the mean free path of strontium below the cell length. Given the observed sharp temperature decrease away from the heated region, the walls quickly act as absorbers and strontium deposition would be hindered. An alternative would be the use of a design similar to \cite{Li2004} to circumvent the problem, enabling arbitrary cell diameter and shorter length.

\section{Measurement accuracy}
\label{subsec:Doppler_shifts}
\label{sec:measurement accuracy}

In this section, we review the main causes for line shifts and distortion  that may alter the accuracy of measurements in both settings. First-order Doppler shifts \cite{Kersten1999} may easily become the dominant effect. We will show that they affect differently the atomic beam and hot vapor configurations, and conclude the section by a brief overview of the weaker effects.

First-order Doppler shifts can arise as a consequence of a deviation from parallelism between the incoming and retro-reflected spectroscopy beams \footnote{It should be noted that the finite transit time broadening, induced by the finite waist of the spectroscopy beams, can also be interpreted in terms of Doppler shifts from the various plane wave components of the Gaussian spectroscopy beams \cite{Borde1976}.}.
The atoms that contribute to the sub-Doppler spectral feature propagate along the wavevector difference of the two spectroscopy beams (orthogonally to the spectroscopy beams if the latter are perfectly parallel). We denote $\alpha$ the angle between the two beams.  For a small angular deviation from perfect retro-reflection, $(\alpha - \pi) \ll 1$, the center of the line for atoms belonging to a single velocity class will be shifted by $\delta \approx (\alpha-\pi) \, v \, k_{\rm{S}} / 2$, where $v$ is the velocity along the wavevector difference. Each velocity class is affected by a different Doppler shift $\delta$, which results in a distortion of the spectrum. 

The consequences then differ for the atomic beam and hot vapor cell geometries, because the velocity distributions are not identical. In the atomic beam setup, the atoms affected by the spectroscopy light are distributed according to $n_2(v)$ defined in section \ref{subsec:atomic beam}. Thus, at a temperature $T = \SI{480}{\celsius}$, a mis-parallelism of $\alpha = \SI{1}{\milli\radian}$ would distort  the line asymmetrically, increasing the FWHM by \SI{310}{\kHz} and shifting its maximum by \SI{190}{\kHz}. By retro-reflecting the spectroscopy beam with a trihedral prism with a maximum beam deviation of \SI{3}{\arcsecond} (\SI{5e-2}{\milli\radian}), we limit the bias below \SI{10}{\kHz} and the broadening below \SI{15}{\kHz}, both of the order of the natural linewidth of the intercombination line.

Interestingly, if the velocity distribution is isotropic in the interaction region -- as we expect in the hot vapor cell -- no shift of the line center occurs. The velocity distribution is separable: $n(v) \propto \exp(-m v^2 / 2 k_B T)$ along any axis. As compared to the directed atomic beam, at the same temperature $T = \SI{480}{\celsius}$, the broadening is increased (22\,kHz FWHM for \SI{5e-2}{\milli\radian}), but the lineshape remains symmetric and no shift of the maximum occurs.

A deviation from parallelism between the readout and spectroscopy beams has a smaller impact. Mostly, as the readout beam interrogates a different set of velocity classes, the sub-Doppler feature shelves a lower number of atoms which results in a lower signal.
Additionally, the Doppler-broadened pedestal on which the sub-Doppler resonance sits can also be shifted for a large misalignment.
Given the large ratio between the widths of the sub-Doppler ($\sim\!\SI{100}{\kHz}$) and Doppler ($\sim\!\SI{20}{\MHz}$) features, this should have a small effect ($<\!\SI{100}{\Hz}$) as soon as the mis-parallelism is less than \SI{1}{\milli\radian}.

We finally give a brief overview of the other weaker contributions to the line shift and distortion, which are common to both setups. The recoil doublet splitting \cite{Borde1979} is $\pm$4.8 kHz. The second-order Doppler shift $-f_0 v^2 / 2 c^2$ \cite{Mungall1971}, with $f_0$ the transition frequency and $c$ the speed of light, is of order  \SI{260}{\Hz}  at the characteristic velocity $v = \bar v_2$, i.e. a $\sim 6\times10^{-13}$ relative shift. For our magnetic fields below 4\,G, the quadratic Zeeman shift is below 10 Hz. The differential AC-stark shift of the intercombination transition from the spectroscopy light is negligible, of order 0.2 Hz for $I \sim \SI{100}{\Isat}$.
Finally, in the hot vapor cell geometry, the overlap of the readout and spectroscopy beams may cause an additional distortion by AC-stark shift at 461\,nm. We did not observe this, probably as the dynamics on the intercombination transition freezes under the influence of readout beam scattering very far from the beam axis. Nevertheless, for an application with high accuracy requirements, a hollow spectroscopy beam \cite{Manek1998, Liu2007} could be advised to prevent the overlap.

\section{Conclusion and perspectives}

In this paper, we have demonstrated the applicability of shelving detection to the spectroscopy of the strontium intercombination line. The method dramatically enhances the signal as compared to direct saturation spectroscopy on the narrow line, which is the present standard on strontium experiments \cite{Li2004}.
Strong signal and low measurement instabilities are then obtained with minimal work on the detection and at low atomic densities, which guarantees a long atomic source lifetime and low power consumption.
Most notably, we have demonstrated the shelving spectroscopy of strontium both in a directed atomic beam and in a hot vapor cell. It could thus be easily and directly implemented in most existing strontium experiments.

On our atomic beam, we evaluated a measurement instability  of $2\times 10^{-12}$ at 1\,s despite working at a low density. This instability is a factor of ten above our estimate of the limit imposed by fundamental shot noises for the experimental parameters. Furthermore, we expect that a second order of magnitude could also be gained from increasing the atomic density, such that the relative instability may be brought down to $3\times10^{-14}$ without any change in geometry. 
Beyond optimizing the dimensions of the spectroscopy and readout beams, more sophisticated improvements have been developed elsewhere that could be applied in the beam geometry, e.g. interrogating with the readout beam only atoms with low Doppler shift on the spectroscopy beam \cite{McFerran2009}, or implementing a Ramsey interferometry scheme with a selected longitudinal velocity class \cite{Kai-Kai2006}. 

Our results naturally highlight the possible application of shelving spectroscopy of the strontium intercombination line to time and frequency metrology. This is emphasized by the strong similarities in frequency- and time-scales with thermal calcium beam clocks\,\cite{McFerran2009, Shang2017}. Here as well, shelving detection should bring photon shot noise behind atomic shot noise \cite{Kai-Kai2006}, which itself is very small due to the large accessible density and continuous operation \cite{Olson2019}. 
The high mass of strontium and moderate temperature requirement mean that a large interrogation time, i.e. a small transit broadening, can be achieved with a small interrogation distance, and possibly without Ramsey techniques. 
Overall, an application as transportable, low-complexity atomic clocks could thus be foreseen. In this work, we optimized more the atomic beam setup; the alternative hot vapor cell approach could be advantageous with regards to compactness, technical simplicity, and source lifetime, as it needs only be heated to less than \SI{400}{\celsius} to maximize the signal or signal-to-noise ratio. It should furthermore offer vanishing first-order Doppler bias thanks to the isotropic atomic velocity distribution.

\begin{acknowledgments}
We acknowledge experimental contributions to this project from Florence Nogrette and Olivier Lopez, and critical reading of the manuscript by Anne Amy-Klein, Benoit Darqui\'e, and Olivier Gorceix. The research performed at LPL was funded by the Agence Nationale de la Recherche (project ANR-16-TERC-0015-01), the Conseil R\'egional d’Ile-de-France, Institut Francilien des Atomes Froids, DIM Nano'K (projects METROSPIN and ACOST), DIM Sirteq (project SureSpin), and Labex FIRST-TF (project CUSAS). The research performed at LCF was funded by the European Research Council under the EU H2020 research and innovation programme (Grant agreement No. 679408--DYNAMIQS). Marc Cheneau acknowledges additional funding from the EU H2020 Fet Proactive Grant No. 640378 (RYSQ) and from the International Balzan Prize foundation through the 2013 prize awarded to Alain Aspect.
\end{acknowledgments}

\appendix

\section{Model of the signal-to-noise ratio and frequency instability}
\label{sec:SNR}

In this appendix we relate the lock-in amplifier signal to physical quantities, to give a qualitative understanding of optimal parameter ranges and fundamental noise limits. We will illustrate how a shelved readout enhances the photonic signal to a point where in principle atomic shot noise draws the limit \cite{Shang2017}, unlike direct saturated spectroscopy of the narrow line where photonic shot noise instead should be the ultimate limiting factor \cite{Ferrari2003, Courtillot2005, Hui2015}. 

This discussion will rely on the assumption that the broad distribution of longitudinal velocities and thus of interaction times smears away the effect of atomic coherences in our spectra; thus, the $\ce{^3P1}$ populations of the atoms passing through the spectroscopy laser reach a stationary state, down to 0 far from resonance and up to 1/2 at resonance. 

We first illustrate general scalings on the frequency instability of the measurement in the limit of low readout beam absorption. Then, we present numerical expectations beyond this assumption. 

\subsection{Approximate scalings}
We trace the signal source back to the readout light intensity transmitted through the atoms, impinging on the photodiode. \textbf{In the limit of small absorption}, as e.g. on the atomic beam data presented here, the number of absorbed photons is roughly proportional to the number $N^{(0)}$ of atoms with transverse velocity $\lesssim \Gamma/k_{\rm{R}}$ in the readout beam at a given time. Here, $k_{\rm{R}} = 2 \pi / \lambda_{\rm{R}}$ is the wavevector of light resonant with the readout transition, while $k_{\rm{S}}$ is the wavevector of light resonant with the spectroscopy transition.
When the saturating spectroscopy light approaches resonance, two transverse velocity classes interact with it and the excited state population of these classes is of order 1/2. \textbf{We assume that the width $\delta v$ of these velocity classes is limited by power broadening on the spectroscopy transition}: $\delta v \simeq \sqrt{2} \Omega / k_{\rm{S}}$.
Then, the number of atoms shelved in $\ce{^3P1}$ from the readout light  is $N_{\rm{shelved}}^{(1)}  \propto N^{(0)} \sqrt{2} \Omega/k_{\rm{S}}$. On the sub-Doppler feature, only a single velocity class is shelved, but with a Rabi frequency  stronger by a factor $\sqrt{2}$. There, the number of shelved atoms is $N_{\rm{shelved}}^{(2)} \propto N^{(0)}  \Omega/k_{\rm{S}} $. In the limit of small absorption, the sub-Doppler signal for the spectroscopy transition, $S$, is proportional to the difference between these shelved atom numbers. At resonance we thus have:
\begin{equation}
 S  \propto N_{\rm{shelved}}^{(1)} - N_{\rm{shelved}}^{(2)} \propto \Omega .
\end{equation}

Assuming that the modulation amplitude $2 \pi \Delta f$ is smaller than the power-broadened linewidth $\gamma(I) \simeq \sqrt{2} \Omega$, the dispersive feature $S'$ produced by the lock-in amplifier 
is proportional to the derivative of $S$ with respect to the spectroscopy laser frequency, multiplied by $\Delta  f$. As both the peak value of $S$ and the spectrum width are proportional to $\Omega$, the extrema of the derivative $S'$ do not depend on $\Omega$, but only on $\Delta f$:
\begin{equation}
 {\rm{max}}(S') \propto \Delta f .
\end{equation}
The slope of $S'$ at resonance then is
\begin{equation}
 S'' \propto \Delta f / \Omega .
\end{equation}
Finally, the single shot frequency measurement instability $\delta f$ is evaluated from the ratio of the noise on $S'$, labeled $\delta S'$, to the slope $S''$.
Under our assumptions, we then have:
\begin{equation}
 \delta f = \frac{\delta S' }{ S''}
 \propto  \Omega \frac{ \delta S' }{ \Delta f}.
\end{equation}

The noise $\delta S'$ can be dominated by technical sources or have a fundamental origin.
\textbf{If the noise $\delta S'$ is technical}, additive and independent of $\Omega$ and $\Delta f$, then the short term instability can be kept optimal and independent of the spectroscopy power if one scales the modulation amplitude $\Delta f$ linearly with the power broadening $\Omega$ (keeping $2 \pi \Delta f < \gamma(I) \simeq \sqrt{2}\Omega$, the prior assumption). 
This corresponds to the discussion of the frequency instabilities in \cref{sec:precision}, \cref{fig:width/slope} and \cref{fig:uncertainty}.

\subsection{Numerical model}

\begin{figure}
 \centering
 \includegraphics[width=1 \columnwidth]{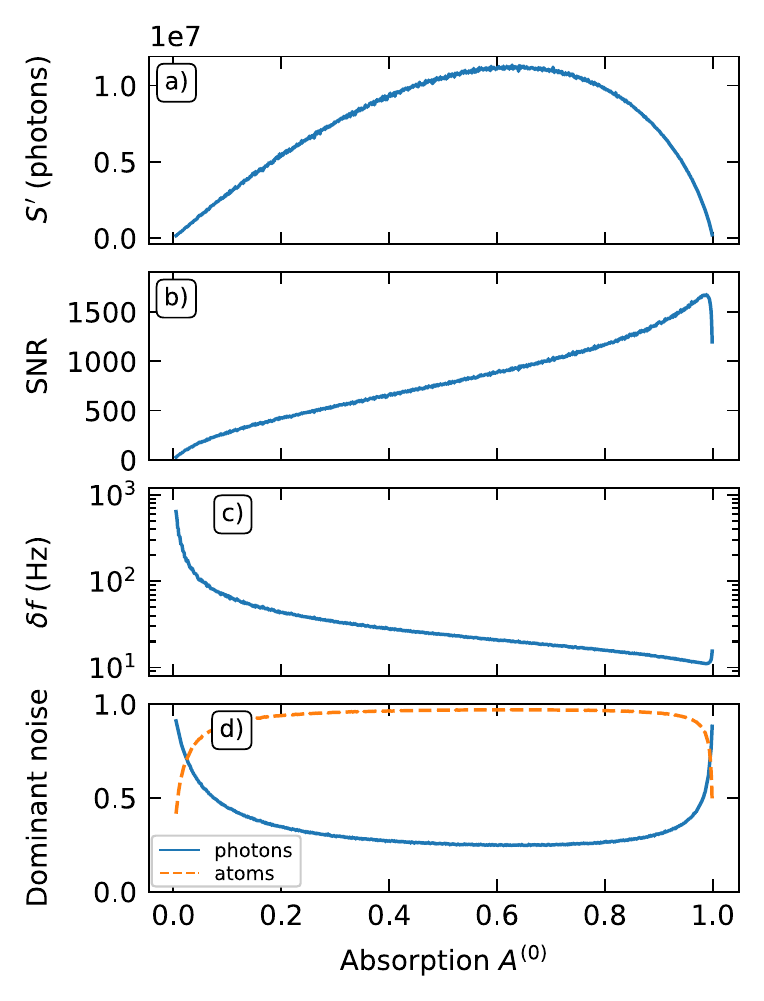}
 \caption{Model of the shot noise limits to the signal-to-noise ratio as a function of the absorption of the readout beam, when the spectroscopy beam is off. The technical parameters (geometry, beam intensities...) correspond to those of the atomic beam spectrum presented in Fig.\,\ref{fig:spectra}, at \SI{83}{\Isat}. a) Signal $S'$, in photon number difference per modulation time. This is proportional to the SNR in case of dominant additive \textit{technical} noise. b) SNR in case of dominant \textit{fundamental} atomic and photonic shot noise. c) Corresponding fundamental limit to the frequency instability at 1\,s (log-lin scale). d) Relative role of the two shot noises, normalized together quadratically to unity.
 }
 \label{fig:SNRmodelling}
\end{figure}

We now develop a numerical model of the signal strength and of the maximum signal-to-noise ratio allowed by fundamental shot noises, that can then be directly related to a minimum frequency instability. We will consider two fundamental sources of noise: atomic shot noise and photonic shot noise on the detected readout light. The model does not assume a low absorption, but \textbf{remains under the assumption of power-dominated broadening}. Its results, under the experimental conditions of our atomic beam experiment, are shown in Fig.\,\ref{fig:SNRmodelling}.

We characterize the readout beam absorption by three values: i) its asymptotic value when the spectroscopy beams are off-resonant, $A^{(0)}$; ii) its value when the spectroscopy beams are close from resonance, but not yet on the sub-Doppler feature, $A^{(1)}$; iii) its value on the sub-Doppler peak, $A^{(2)}$.
Connecting absorption and optical density, i.e. $A^{(i)} = 1- \exp(-OD^{(i)}), i \in \{0,1,2\}$, we now describe the shelving effect. 
The ensemble of $N^{(0)}$ atoms responsible for the absorption of the readout beam  has a Doppler-limited velocity spread along the readout beam axis: $\vec k_{\rm{R}} \cdot \vec \delta v \simeq \Gamma$. The atomic ensemble shelved by a single spectroscopy beam is restrained in velocity spread to $\vec k_{\rm{S}} \cdot \vec \delta v_{\rm{S}} \simeq \sqrt{2}\Omega$. With two beams, there are either two such classes shelved (when the spectroscopy light is off-resonance), or one class only (when on resonance) but with effective Rabi frequency stronger by a factor $\sqrt{2}$. We approximate the population in the shelved state $\ce{^3P1}$ to be of order 1/2 for atoms resonant with at least one spectroscopy beam. Then, accounting for the shelved atoms, the readout optical density away from and on the sub-Doppler line are respectively:
\begin{eqnarray}
 OD^{(1)} &=& OD^{(0)} \left( 1 - \frac{\sqrt{2} \Omega}{\Gamma} \frac{\lambda_{\rm{S}}}{\lambda_{\rm{R}}}   \right) 
 \\
 OD^{(2)} &=& OD^{(0)} \left( 1 - \frac{\Omega}{\Gamma} \frac{\lambda_{\rm{S}}}{\lambda_{\rm{R}}}  \right) 
\end{eqnarray}

We now construct the lock-in amplifier peak signal ${\rm{max}}(S')$, that we will write for simplicity $S'$. We assume a low frequency modulation ($2500\,
\rm{Hz}< \gamma/2\pi$). For an optimal amplitude of the frequency modulation, comparable to the linewidth, 
we approximate $S'$ by a measurement of the difference in transmitted readout beam intensities when the spectroscopy beams are on- and off-resonance with the sub-Doppler feature:
\begin{eqnarray}
 S' &=& G \left( P \tau \frac{ \lambda_{\rm{R}}}{h c} (1-A^{(2)}) \right)
 -
 G \left( P \tau \frac{ \lambda_{\rm{R}}}{h c} (1-A^{(1)}) \right) \nonumber
\\
&=& G \left(N_{ph}^{t,(2)} - N_{ph}^{t,(1)}\right)
 .
 \label{eqqqq}
\end{eqnarray}
Here, the terms in parenthesis are photon numbers, transmitted to the photodiode. $P$ is the readout beam power and $\tau$ a characteristic collection time. G includes detection gain parameters from the photodiode to the lock-in amplifier output, that we assume noiseless.

We treat the measurements of $N_{ph}^{t,(2)}$ and $N_{ph}^{t,(1)}$ as independent. The shortest timescale is the mean transit time of the atoms in the readout beam, $\tau = t_{\rm{sample}}$, after which the atomic sample is renewed. We approximate the Gaussian beam with waist $w$ along the atomic beam direction by a uniform beam with radius $w/\sqrt{2}$. Then, $t_{\rm{sample}} \simeq (\sqrt{\pi/2}\cdot w) / \bar v_2 \simeq \SI{1.7}{\us}$ on the atomic beam setup. 
To estimate the effects of atomic and photonic noise on the dispersive lock-in amplifier signal $S'$, we reason as follows. 
\begin{enumerate}
\item For each of the measurements of $N_{ph}^{t,(2)}$ and $N_{ph}^{t,(1)}$, we make the following estimates: 
\begin{itemize}
\item 
before absorption by the sample, the incident photon number is $\langle N_{ph}^i \rangle =  \frac{P \tau \lambda_{\rm{R}}}{h c}$. 
It is a Poissonian process. 
The transmission of these photons is a random process for each photon, with probability $T = (1-A)$.

\item As the atomic sample is renewed every $t_{\rm{sample}}$, the number of atoms in the readout beam fluctuates, and the transmission probability $T$ for each incident photon also fluctuates : $\langle T^2\rangle   \neq \langle T\rangle^2$, where averages run over the successive realizations of the atomic sample.

We estimate the average number of atoms interrogated at a given time from the optical densities and beam waists ($w$ along the atomic beam, $w_\perp$ transverse to it): 
\begin{eqnarray}
\langle N^{(0)} \rangle & \simeq & \frac{\pi  w w_\perp }{2} \cdot  \rm{OD} \cdot \frac{2 \pi }{ 3 \lambda_{\rm{R}}^2}, \\
\langle N^{(1)} \rangle & = &  \langle N^{(0)}\rangle \left( 1 - \frac{\sqrt{2} \Omega}{\Gamma} \frac{\lambda_{\rm{S}}}{\lambda_{\rm{R}}}   \right), \\
\langle N^{(2)} \rangle & = &  \langle N^{(0)}\rangle \left( 1 - \frac{\Omega}{\Gamma} \frac{\lambda_{\rm{S}}}{\lambda_{\rm{R}}}   \right). 
\end{eqnarray}

Assuming Poissonian fluctuations around these mean atom numbers, we estimate the variance of the transmission coefficient ${\rm{Var}}(T) = \langle T^2\rangle - \langle T\rangle^2$ by sampling the Beer-Lambert absorption law.

\item We now compute the statistics of the photon number transmitted to the photodiode during $t_{\rm{sample}}$. Combining the Poissonian statistics of the incident photon number $ N_{ph}^i$ and the fluctuating transmission coefficient $T$, we can demonstrate that the transmitted photon number has the statistics:
\begin{eqnarray}
\langle N_{ph}^t \rangle &=& \langle T \rangle \langle   N_{ph}^i \rangle  \\
{\rm{Var}}(N_{ph}^t) &=& \langle T \rangle \langle  N_{ph}^i \rangle + {\rm{Var}}(T) \langle N_{ph}^i \rangle^2 
\label{eq:fluctuat}
\end{eqnarray} 
In Eq.\,\ref{eq:fluctuat}, the term $\langle T \rangle \langle  N_{ph}^i \rangle$  is Poissonian photon shot noise in the transmitted light, and the term ${\rm{Var}}(T) \langle N_{ph}^i \rangle^2$ encompasses the effects of the fluctuating transmission coefficient, due to atomic shot noise. 
This formula could easily be extended to include technical fluctuations of the readout laser intensity.

\end{itemize}
\item We now combine the effect of the differential measurement on resonance and away from resonance,  $S' = G (N_{ph}^{t,(2)} - N_{ph}^{t,(1)})$ treated as two independent measurements, and the effect of time integration from the timescale $t_{\rm{sample}}$ up to the lock-in amplifier integration time $t_{\rm{integr}}$. We account for the fact that the total measurement time is divided between the terms (1) and (2). Then, the signal-to-noise ratio, limited by shot noises, is:
\begin{equation}
\rm{SNR} = \sqrt{\frac{t_{\rm{integr}}}{2\, t_{\rm{sample}}}}
\cdot \sqrt{\frac{\left(\langle N_{ph}^{t, (1)} \rangle -  \langle N_{ph}^{t, (2)} \rangle\right)^2}{{\rm{Var}}(N_{ph}^{t, (1)}) + {\rm{Var}}(N_{ph}^{t,(2)})}}
\end{equation}
\end{enumerate}
For a Lorentzian line, the frequency measurement instability $\delta f$ at the timescale $t_{\rm{integr}}$  is related to the SNR by
\begin{equation}
\delta f = \frac{\gamma(I)}{2 \pi} \frac{3 \sqrt{3}}{32} \frac{1}{\rm{SNR}}
\end{equation}
where $\frac{\gamma(I)}{2 \pi}$ is the line FWHM including all broadening mechanisms: $\frac{\gamma(I)}{2 \pi} \simeq 114$\,Hz in the conditions of Fig.\,\ref{fig:spectra}.

The result is shown in \cref{fig:SNRmodelling}, varying the absorption but otherwise in the geometry and conditions of the atomic beam spectrum of Fig.\,\ref{fig:spectra}. Thanks to the shelving detection on a broad line, the effect of photonic shot noise is below the effect of atomic shot noise for absorptions from 0.06 to 0.98.
We observe that the maximal signal would be obtained for an absorption of order 0.6. The fundamental limit to the SNR is obtained for even higher absorption. In the regime of our experiments on the atomic beam (7$\%$ absorption), the fundamental limit to the SNR is 220, about a factor of 10 higher than ours. By going to a very strongly absorbing medium, the fundamental limit to SNR could reach 1500, corresponding to a frequency instability at 1\,s  down to 12\,Hz, i.e. a relative frequency instability of $3\times10^{-14}$.

\section{Robustness against a secondary reference}
\label{subsec:reproducibility}

\begin{figure}
 \centering
 \includegraphics{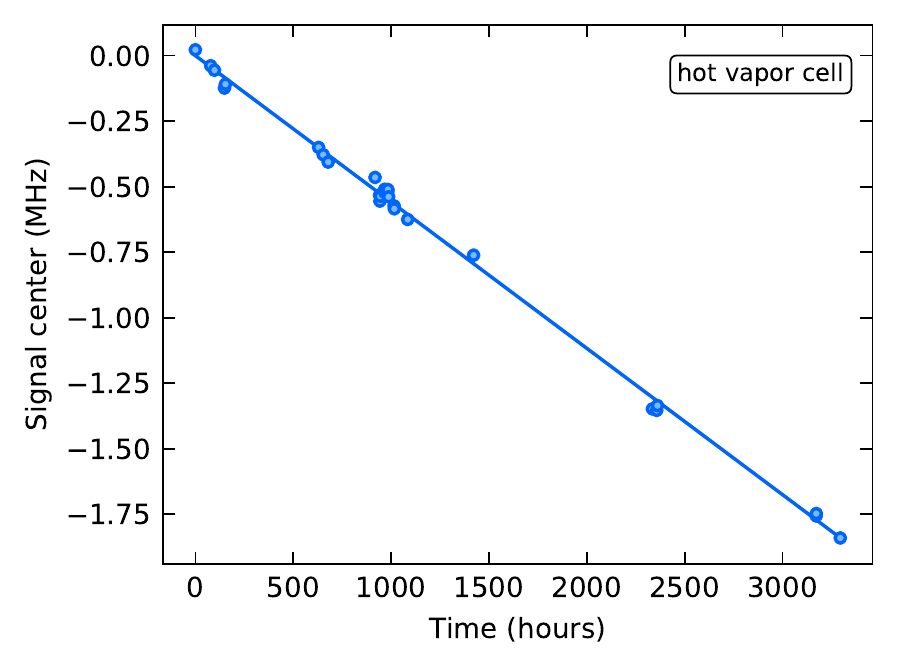}
 \caption{
  Line center position relative to the reference optical cavity as a function of the time at which the measurement was performed.
  29 measurements performed with the hot vapor cell are gathered in this plot, corresponding to a wide variety of experimental settings: spectroscopy beam power, modulation amplitude, etc.
  The measured data (circles) closely follows a linear regression (solid line) with a slope of \SI{-155.11 \pm 0.5}{\mHz\per\s}.
 }
 \label{fig:cavity_drift}
\end{figure}

A complete characterization of a frequency reference involves a comparison to a secondary reference. In the absence of an external absolute reference, we measure the frequency difference between the shelving spectroscopy line and the resonance frequency of the ultra-stable optical cavities used to stabilize our spectroscopy lasers. This measurement, at short and long term, provide very preliminary insight in the robustness of the shelving spectroscopy and in the long-term behavior of the ultrastable cavities. Both setups rely on commercial ULE optical cavities, temperature stabilized, under vacuum (Stable Laser Systems).

On the atomic beam setup, on a short timescale of one day, we characterized the measurement robustness from data (including those of Figs.\,\ref{fig:spectra} - \ref{fig:uncertainty}) acquired with a wide range of spectroscopy beam intensity (20 to 400\,$I_{\rm{sat}}$) and modulation amplitude (12.5 to 115\,kHz). A linear drift with time of the line centers is measured; the standard deviation of the measurements around this drift was of \SI{1.6}{\kHz}, providing an upper bound for the general robustness of the measurement. The strongest deviations, of order 10 kHz, all arose from measurements with large spectroscopy power $\geq$\SI{180}{\Isat}. 

With the hot vapor cell setup, we have performed measurements over a longer time span of 6 months. The result is summarized in \Cref{fig:cavity_drift}. These data were recorded using a similarly broad range of experimental parameters (spectroscopy beam power, modulation amplitude, etc.). 
The cavity remained at constant temperature over the entire time span. The main feature is a linear frequency drift between the two references, with a fitted slope of \SI{-155.11 \pm 0.5}{\mHz\per\s}.
Such a drift is expected when using an optical cavity as frequency reference, and may be attributed to the aging of the cavity spacer material \cite{Dube2009, Hagemann2014}. The r.m.s. value of the fit residuals is \SI{24}{\kHz} over 6 months; we cannot discriminate between the robustness of the shelving spectroscopy measurement and fluctuations of the cavity resonance that would deviate from the linear drift.

\end{document}